\documentclass[12pt,a4]{article}
\usepackage{epsfig}
\input{epsf}
\Roman{table}

\newcommand{\sh}{\mbox{sh}}
\newcommand{\ch}{\mbox{ch}}

\begin{document}

\title{Dispersion Relation in a Ferrofluid Layer of Any Thickness  and Viscosity in a Normal Magnetic Field; Asymptotic Regimes }
\bigskip
\author{B. Abou $^1$, G. N\'eron de Surgy $^2$ and J. E. Wesfreid $^1$}

\date{}
\setcounter{page}{1}
\maketitle
\bigskip
{\footnotesize
$^1$ Laboratoire de Physique et M\'ecanique des Milieux H\'et\'erog\`enes (PMMH-Unit\'e~CNRS~857), Ecole Sup\'erieure de Physique et Chimie Industrielles de Paris (ESPCI), 10 rue Vauquelin, 75231 Paris Cedex 05, France. E-mail: abou@pmmh.espci.fr\\
$^2$ Laboratoire de G\'enie Electrique de Paris (LGEP), Plateau du Moulon, 91192 Gif sur Yvette Cedex, France.\\

\bigskip
Version abr\'eg\'ee du titre: Dispersion relation in ferrofluid
layers.\\


\newpage

\begin{itemize}

\item Titre en fran\c{c}ais: Relation de dispersion d'une couche de ferrofluide de viscosit\'e et d'\'epaisseur quelconques sous champ magn\'etique normal; R\'egimes asymptotiques.
 
\item Classification Physics Abstracts:
 \begin{enumerate}
  \item 47.20. Hydrodynamic stability and instability.
  \item 75.50.Mm. Magnetic liquids.
  \item 47.35. Waves.
 \end{enumerate}


\item {\bf{ R\'esum\'e en Fran\c{c}ais:}}

Nous avons calcul\'e la relation de dispersion des ondes de surface
dans une couche de ferrofluide d'\'epaisseur et de viscosit\'e
quelconques, soumis \`a un champ magn\'etique normal \`a sa surface
(instabilit\'e de pics de Rosensweig). Cette relation montre que le
champ magn\'etique critique et le vecteur d'onde critique de
l'instabilit\'e d\'ependent de l'\'epaisseur de la couche de
fluide. La relation de dispersion a \'et\'e simplifi\'ee pour quatre
r\'egimes asymptotiques: couche \'epaisse ou mince et comportement
visqueux ou inertiel. Nous avons calcul\'e les valeurs critiques de
l'instabilit\'e dans ces quatre cas. Nous montrons qu'un param\`etre
typique du ferrofluide permet de savoir dans quel r\'egime, visqueux
ou inertiel, se situe le ferrofluide pr\`es du seuil de l'instabilit\'e.

\end{itemize}
\newpage

\begin{abstract}
We have calculated the general dispersion relationship for surface waves
on a ferrofluid layer of any thickness and viscosity, under the influence of a
uniform vertical magnetic field. The amplification of these waves can induce an
instability called peaks instability (Rosensweig instability). The expression of the dispersion
relationship requires that the critical magnetic field and the
critical wavenumber of the instability depend on the thickness of the
ferrofluid layer. The dispersion relationship has been simplified into
four asymptotic regimes: thick or thin layer and viscous or inertial
behaviour. The corresponding critical values are presented. We show that a typical
parameter of the ferrofluid enables one to know in which regime, viscous
or inertial, the ferrofluid  will be near the onset of instability.

\end{abstract}

\section{Introduction}
\label{Introduction}

It is known since the experiment performed by Cowley and Rosensweig
(\cite{rosens}, \cite{rosensweig}) that a normal magnetic field has a
destabilizing influence on a flat interface between a magnetizable
fluid and a non magnetic one. Above the magnetic induction threshold
$H_{crit}$, the initially flat interface exhibits a stationary
hexagonal pattern of peaks (Rosensweig instability). The
hexagonal pattern becomes a square pattern above another critical magnetic induction $H_{crit}'>H_{crit}$ \cite{gailitis}.

In this paper, we want to obtain a general dispersion equation
(ferrofluid layer of any thickness and any viscosity under the
influence of a uniform vertical magnetic field) by linearizing the
equations governing the problem. This allows to calculate the critical
values of the instability: wavenumber and magnetic
induction. Nevertheless, linearized equations are inadequate for the
full description of either phenomenon: the symmetry of the developed
wave array (hexagonal or square) and the wave amplitudes may be
determined only from nonlinear equations.  It has been treated by
Gailitis in 1977 \cite{gailitis}. A generalized Swift-Hohenberg
equation constitutes a minimal model to account for the formation of
hexagons as well as squares. In such an envelope equation, for which
the quadratic term is sufficiently small, hexagons are to be expected
from the modeling \cite{herrero}. Furthermore, a weakly nonlinear
analysis involves that the dynamics of the patterns depends on their symmetry \cite{brand-wesfreid}.\\
We shall remember previous calculations of dispersion equations in
section (\ref{previous}) and section (\ref{elec}) will put in evidence
the analogy with the electric case. Indeed, a layer of liquid metal
under a normal electric field develops a similar instability of peaks
(Taylor cones \cite{taylor}). We applied to the magnetic case the
previous analyse of Hynes \cite{hynes} and Limat \cite{limat} for the
gravitational amplification of capillary waves (Rayleigh-Taylor
instability) and of N\'eron de Surgy et al. \cite{neron} for the
electric amplification of capillary waves (Taylor cones)
\cite{taylor}. The general dispersion relationship that we obtain in
section (\ref{equation}) leads to the fact that the magnetic induction
threshold and the critical wavenumber depend on the thickness of the
ferrofluid layer \cite{brand}. We derive the asymptotic behaviour in
various regimes and give analytical dispersion equations in the case of
thick or thin, inertial or viscous layers. From a general approach, we then recover three
previously known results (thick-inertial \cite{rosensweig}, thick-viscous
\cite{salin}, thin-viscous \cite{bac88}) obtained from various
approaches. Furthermore, we give an explicit expression of the critical
magnetic induction in the case of a thin layer of ferrofluid: it
differs from the one of the thick case. We also demonstrate the influence of the ratio
$l_v/l_c$, where $l_c$ is the capillary length and $l_v$ the viscous
length of the ferrofluid: depending on the value of this ratio, the
ferrofluid will have a viscous or inertial behaviour near the onset of the instability.

\subsection{Previous approaches}
\label{previous}
Ferrofluids or magnetic liquids are permanent, colloidal suspensions
of ferromagnetic particles ($100$ ~\AA) in various carrier
solvents. Fluid instabilities can arise with these liquids, especially
surface instabilities. In the case of the peaks instability, the study
of the linear stability of the interface, by Cowley and Rosensweig
\cite{rosens}, between an half-infinite and inviscid ferrofluid of
density $\rho$ and the vacuum, with normal modes of perturbation, leads to the following dispersion relation:

\begin{equation}
\label{ros}
\rho s^2=-\rho gk-\gamma k^3+\frac{\mu_0(\mu/\mu_0-1)^2}{1+\mu_0/\mu}H_0^2k^2
\end{equation}
where $s$ is the growth rate of perturbation, $k$ (the modulus of
${\bf{k}}$) is the horizontal wavenumber, $\mu$ is the magnetic
permeability of the ferrofluid, $\gamma$ is the interfacial tension
ferrofluid-vacuum, $H_0$ the modulus of the normally applied magnetic
induction and $g$ the gravitational field.\\
The critical values of the instability are then:
\begin{equation}
\label{critepais}
\left\{
\begin{array}{lrc}
H_{crit}=\sqrt{\frac{2}{\mu_0}\frac{(1+\mu_0/\mu)}{(\mu/\mu_0-1)^2}\sqrt{\rho g \gamma}}\\
k_{crit}= \sqrt{\rho g/\gamma}=k_c=1/l_c
\end{array}
\right.
\end{equation}
where $k_c$ is the capillary wavenumber.\\
Above the threshold of linear stability, a broad band of wavenumber is
theoretically found unstable. It is seen from other instabilities that
the mechanisms of wavenumber selection are very complex, including the
effects of sidewalls, time-dependent effects in cellular structures,
axisymmetric stucture in the case of Eutectic solidification, the nonlinearities of the problem, etc..\cite{wesfreid-zaleski}. A common choice has been to consider
then that one wavenumber $k_m$ is selected. This wavenumber
corresponds to the maximum growth rate $s_m$ of the small undulations,
solution to the following equation $\partial_k s|_{H=const}=0$. This
prediction has only be
checked when $H_0$ increases very quickly from a subcritical value to
$H_{crit}$. In most cases, the wavenumber remains equal to $k_{crit}$
above the onset, whatever the symmetry is or the experimental
procedure used to increase the field (continuous increase
\cite{rosens}, field jumps \cite{allais} or alternating
field \cite{dortona} at different frequencies). This encouraged
D. Salin, in 1993, to take into account the effect of the ferrofluid
viscosity, adapting the Landau and Lifshitz approach \cite{landau}. His result leads to a wavenumber $k_m$ equal to $k_{crit}$,
whatever the field is \cite{salin}. Simultaneously, N\'eron de Surgy et
al. \cite{neron} obtain the same result for metal liquids. As seen above,
the experimentally selected wavenumber is not obviously $k_m$ nor $k_{crit}$ \cite{abou}. We shall see in section
(\ref{numerical}) that the description of the unstable behaviour of
the ferrofluid (viscous or not) near the onset can be determined by
the value of the ratio $l_v/l_c$. When this ratio approaches zero, an
inertial description is actually sufficient. 

\subsection{Analogy with electro-capillary instability}
\label{elec}

In 1993, G. N\'eron de Surgy et al. studied the linear growth of
electro-capillary instabilities in the very general case where the
viscosity of the fluid and its thickness are of any value
\cite{neron}, following the studies of Hynes \cite{hynes} and Limat
\cite{limat}. The study of this instability is similar to our study:
an electric field that is applied normally to the free surface of a
conducting fluid (mercury, for example) has a destabilizing effect on
this interface. In their paper, they derived the asymptotic behaviour
in various regimes and gave dispersion equations in the case of thin
or thick, inviscid or viscous films: we present a similar approach for
the magnetic case (but note that substituting ${\bf{E}}$ for ${\bf{H}}$
in the electric dispersion equations does not result in the correct
magnetic dispersion equations). Furthermore, we note that the critical
electric field and the critical wavenumber of the electro-capillary instability does not depend on
the thickness: While ${\bf{E}}={\bf{0}}$ in the whole layer of
mercury, there is penetration of the magnetic field in the ferrofluid
and the distribution of this field obviously depends on the layer thickness (boundary conditions). 

\section{Characteristic scales for various asymptotic regimes}
\label{scale}

We consider an incompressible and viscous film of magnetic fluid of density $\rho$, dynamic viscosity $\eta$, kinematic viscosity $\nu$ and magnetic permeability $\mu$, that occupies the space between $z=0$ and $z=-a$, in the vacuum. The geometry is supposed to be infinite for both $x$ and $y$ and the medium above and below the ferrofluid is also infinite, so that we consider a two dimensional $(x,z)$ problem (Figure \ref{ferrovide}).\\
If we want to introduce the {\it thickness effects}, we have to consider a horizontal length scale $l_x=\lambda/2\pi$ (with $\lambda = 2\pi /k$), and a vertical length scale $l_z$. The value of $l_z$ will be taken as the lowest value between the film thickness $a$ and $\lambda/2\pi$.\\
For the study of the {\it viscosity effects}, we shall introduce the Reynolds number (Re~=~inertia forces/viscous forces). The Reynolds number is:
$$
\mbox{Re}=\left|\frac{\rho\partial_t v}{\eta \triangle v}\right|\simeq \frac{|s|}{\nu [(1/l_x)^2+(1/l_z)^2]}
$$
where $v$ is the fluid velocity and $\triangle$ the Laplacian
operator.\\
-for a thick film, where the vertical scale is $\lambda/2\pi$, $\mbox{Re}=\frac{|s|}{2\nu k^2}$.\\
-for a thin film, where the vertical scale is $a$, $\mbox{Re}=\frac{|s|a^2}{\nu}$.\\
If $\mbox{Re}\gg 1$, we may neglect viscosity and call the film
inertial, and if $\mbox{Re}\ll 1$, we may neglect inertia
and call the film viscous. \\
This study results in four asymptotic behaviours (Table
\ref{tabregime}, \cite{hynes}, \cite{limat}, \cite{neron}).\\

\section{Equations of the problem}
\label{equation}

We shall present the equations governing the system shown in Figure
(\ref{ferrovide}). Linearity allows us to treat the different Fourier components separately. We shall
denote $\xi(x,z)$ the deformation of the interface, ${\bf{v}}(x,z)$
the velocity of the fluid, ${\bf{H}}={\bf{H_0+h}}$ the perturbed
magnetic induction and ${\bf{n}}$ the unit vector normal to the
interface where ${\bf{n}}\simeq (\partial_x\xi,-1){\bf{u_z}}$ at first
order. We consider $\mu(H)=\mu(H_{crit})$ because our analysis is
valid near the onset (see section (\ref{implicit})).\\
Local equations governing the motion of the magnetic fluid are:
$$
\left\{
\begin{array}{lrc}
{\bf \nabla}\cdot {\bf v}=0 & \mbox{(continuity equation)}\\
\rho [\partial_t {\bf v} + ({\bf v\cdot \nabla}){\bf v}]= -{\bf \nabla}~p + \eta \triangle{\bf v} + \rho {\bf g} & \mbox{(Navier-Stokes equation)}
\end{array}
\right.
$$
Other governing relationships are the Maxwell equations:
$$
\left\{
\begin{array}{lr}
{\bf \nabla} \times {\bf H}={\bf 0} & \mbox{(no charges, no currents)}\\
{\bf \nabla}\cdot {\bf H}=0 
\end{array}
\right.
$$
The boundary conditions (where $[[X]]=~$value of $X$ above the
interface $-$ value of $X$ under the interface) are given by:
$$
\left\{
\begin{array}{lr}
\partial_t \xi= v_z-v_x\partial_x \xi-v_y \partial_y \xi& \mbox{at $z=\xi$,     (free surface condition)}\\
-[[p]]n_i + [[T_{ik} + \sigma '_{ik}]]n_k - (\gamma /R)n_i=0 & \mbox{at $z=\xi$, (sress balance at the interface)}\\
\lbrack\lbrack {\bf n} \cdot \mu {\bf H}\rbrack\rbrack  = 0 & \mbox{at $z=\xi$ and $z=-a$}\\
\lbrack\lbrack {\bf n} \times {\bf H}\rbrack\rbrack  = 0 & \mbox{at $z=\xi$ and $z=-a$}\\
v_z=0& \mbox{at $z=-a$}\\
v_x=0&  \mbox{at $z=-a$}\\
\end{array}
\right.
$$
where $T_{ik}$ is the stress tensor given by (\cite{rosens}, \cite{rosensweig}):
$$
T_{ik}=\mu H_i H_k-\frac{\mu}{2} H^{2} \delta_{ik}
$$
and $\sigma'_{ik}$ is the viscous rate-of-strain tensor given by:
$$
\sigma'_{ik} =\eta (\partial_{x_k} v_i + \partial_{x_i} v_k)
$$
and $R^{-1}$ is the curvature of the interface (positive if directed towards the fluid):
$$
R^{-1}\simeq  -(\frac{\partial^2}{\partial
  x^2}+\frac{\partial^2}{\partial y^2}) \xi
$$
We seek solutions with the following form: 
$$
A=\Re e[\hat{A}(z)exp(st-ikx)]
$$
for $\xi$, $h_x$, $h_z$, $v_x$, $v_z$.
At first order, we get a system of nine linear equations in nine
unknown. In order to find a non trivial solution, the following
determinant has to equal zero. Similar calculations can be read in the electro-capillary case \cite{neron}.

\vspace{1cm}
\centerline{$
\left|
\begin{array}{ccccccccc}
0 & \sh (ka) & 0 & \sh (qa) & 0 & 0 & 0 & 0 & 0\\
k & k \ch (ka) & q & q \ch (qa) & 0 & 0 & 0 & 0 & 0\\
(2\eta k+\frac{\eta s}{\nu k}) \ch (ka) & 2\eta k+\frac{\eta s}{\nu k} 
& 2\eta q \ch (qa) & 2\eta q & i\mu_0 H_0 & \rho g+\gamma k^2 & -i\mu_0 H_0 
& i \mu_0 H_0 e^{ka} & 0\\
2\eta k^2 \sh (ka) & 0 & \eta (k^2+q^2) \sh (qa) & 0 & 0 & 0 & 0 & 0 & 0\\
0 & 0 & 0 & 0 & 0 & 0 & e^{ka} & 1 & -1\\
\sh (ka) & 0 & \sh (qa) & 0 & 0 & -is & 0 & 0 & 0\\
0 & 0 & 0 & 0 & 0 & 0 & e^{ka} & -1 & \mu_0/\mu\\
0 & 0 & 0 & 0 & -1 & -ikH_0(\mu_0/\mu -1) & 1 & e^{ka} & 0\\
0 & 0 & 0 & 0 & \mu_0/\mu & 0 & -1 & e^{ka} & 0
\end{array}
\right|=0 $}
\vspace{1cm}
with $q^2=k^2+s/\nu$.\\
We shall now use dimensionless values with capillary scale as
reference as seen in Table \ref{tabcapillary}.
We introduce the parameter $f=(l_v/l_c)^{3/2}$ (denoted $d$ in
\cite{neron}) where the viscous length $l_v=\nu ^{2/3} g^{-1/3}$
(\cite{hynes}, \cite{limat}). We obtain that $q^2=k^2+\frac{s}{f}$\\
We also denote $\Phi=H_0^2/(H_{crit}^{thick})^2$ where $H_{crit}^{thick}=H_{crit}$ (section (\ref{previous})).
The dispersion relation $s=s(H_0,k)$ is an implicit equation and its dimensionless form can be written:
\begin{eqnarray}
\label{disprel}
4qk^3(q-k\coth(qa)\coth(ka))-(k^2+q^2)^2(q\coth(ka)\coth(qa)-k)+
\frac{4qk^2(k^2+q^2)}{\sh (qa) \sh (ka)}\nonumber\\
=\frac{1}{f^2}(k+k^3-2\Phi k^2 \frac{1+\mu_0/\mu}{1+
\frac{\mu_0/\mu}{1+F(ka)}})(q\coth(qa)-k\coth(ka))
\end{eqnarray}
where 
$$
F(ka)=\frac{(1-\mu/\mu_0)e^{-ka}}{(1+\mu/\mu_0\coth(ka)) \sh(ka)}
$$
We can develop equation (\ref{disprel}) when $s$
tends to $0$ ($s=0$ is a root of equation (\ref{disprel})). We
consider then that $\delta k=q-k\sim \frac{s}{2fk^2}$ and we obtain
that $k+k^3-2\Phi k^2 \frac{1+\mu_0/\mu}{1+
\frac{\mu_0/\mu}{1+F(ka)}}\sim \delta k$. $\delta k=0$ (i.e. $s=0$)
leads to the curve of marginal stability. We obtain that $k_{crit}$
and $\Phi_{crit}$ depend on $a$ (\cite{rosensweig}, \cite{brand}).
We shall now study the dispersion relationship in asymptotic cases.

\section{Asymptotic behaviour of the dispersion relation}
\label{asymptotic}

\subsection{Thick film}

The equation (\ref{disprel}) can be simplified in the regime $a\gg\lambda$ ($ka\gg1$).

\subsubsection{Thick-inertial film}

This corresponds to the case where $Re\gg1$ i.e. $\frac{|s|}{2f
  k^2}\gg1$ (dimensionless Reynolds). \\
The equation (\ref{disprel}) becomes the well-known Cowley and Rosensweig's result \cite{rosens}:

\begin{equation}
s^2=-k^3+2\Phi k^2-k
\end{equation}
where $\Phi_{crit}=1$ and $k_{crit}=1$.
As was shown previously, the wavenumber corresponding to the maximum growth of perturbations, is found to be field-dependent:
$$k_m=\frac{1}{3}(2\Phi+\sqrt{4\Phi^2-3})$$\\
Let us consider the parameter $\varepsilon$ so that
$\Phi=1+\varepsilon$ and the parameter $\delta k$, where $k=1+\delta k$ ($\delta k$ depending on which wavenumber(s) is (are)
actually selected). Near the onset, we develop the growth rate
$s(\Phi, k)$ into a series in $\varepsilon$ and $\delta k$. We found at lowest order in $\varepsilon$ and
$\delta k$ ($\varepsilon$ and $\delta k$ independent): 
$$s^{2}(\Phi=1+\varepsilon,k=k_{crit}+\delta k)=s^{2}(\varepsilon,\delta k)=2\varepsilon-\delta k^2$$
As seen above, a band of wavenumbers of width $\varepsilon^{1/2}$
near threshold is unstable \cite{wesfreid-zaleski}. This means that the order of magnitude $\delta
k^2$ is at most $\varepsilon$ so we can take
$s^2$ to be of order $\varepsilon$.\\
If $k_m$ is actually selected, we check that $\delta k^2$ is of order of
magnitude $\varepsilon^2$ (which is less than $\varepsilon$). If the wavenumber
selected is $k=1+\sqrt{2\varepsilon}$ (we remain on the curve of
marginal stability), we have checked that a similar analysis is
adequate, by making calculations at the following order.
The validity conditions, at the lowest order, for this regime can be
summed up as follows:
$$
\left\{
\begin{array}{lrc}
a\gg1& \mbox{($ka\gg1$, thick film)}\\
f^2\ll\varepsilon & \mbox{($\frac{|s|}{2fk^2}\gg1$ and $\frac{|s|}{2fk^2}$ of order of $\varepsilon^{1/2}/f$, inertial film)}
\end{array}
\right.
$$

\subsubsection{Thick-viscous film}

In this case, $Re\ll1$ i.e. $\frac{|s|}{2f k^2}\ll1$. The equation
(\ref{disprel}) leads to the viscous-dominated relation
(\cite{neron}, \cite{salin}):
\begin{equation}
s=\frac{1}{2f}\left(-k+2\Phi-\frac{1}{k}\right)
\end{equation}
where $\phi_{crit}=1$, $k_{crit}=1$ and $k_m=k_{crit}=1$, whatever
$\Phi$ is.\\
Near the onset of instability, the growth rate becomes after development:
$$
s(\varepsilon,\delta k)=(2\varepsilon-\delta k^2)/(2f) 
$$
We can then take $s$ to be of order $\varepsilon/f$. 
The validity conditions become:
$$
\left\{
\begin{array}{lrc}
a\gg1\\
f^2\gg\varepsilon
\end{array}
\right.
$$
The condition of viscous regime are always true at the onset.

\subsubsection{Range of the asymptotic regimes}
\label{numerical}

As seen above, the cross-over between the inertial and the viscous regime appears when
$\varepsilon=f^2$ where $\varepsilon\ll 1$ (Figure \ref{dthick}). This
relation yields the following consequences:\\
$\bullet$ Strictly at the onset of instability ($\varepsilon=0$), the condition
of inertial regime is never reached: the ferrofluid has a
viscous behaviour ($f$ is finite).\\
$\bullet$ By increasing the magnetic induction, we reach the inertial regime above $\varepsilon=f^2$ (while keeping $\varepsilon\ll 1$).\\
For example, with the ferrofluid APG 512 A ($f\simeq 0.27$),
the inertial regime takes place when
$H_0\simeq H_{crit}^{thick}(1+\frac{\varepsilon}{2})=H_{crit}^{thick}(1+\frac{f^2}{2})$
i.e. $H_0\simeq 1.04~H_{crit}^{thick}$.\\
In Table \ref{thick-range}, we present different values of the parameter $f$.
With the ferrofluid EMG 507, the inertial regime is reached when
$H_0\simeq 1.00002~H_{crit}^{thick}$: we can then consider that the
description thick-inertial is adequate, whatever $H_0$ is.\\
It appears then, that the regime viscous or inertial, near the onset, depend on the type of ferrofluid. One also could use a
ferrofluid, for example APG 067 or APG 314, that enables to remain in
the viscous regime far enough from the onset.  Physical data of
various ferrofluids are presented in Table \ref{data}.\\

\subsection{Thin film}

It seems obvious that the thick case is easier to experiment than the
thin case. In order to stick to an opinion, the typical ferrofluid APG 512 A has a capillary length $l_c\simeq 1.7$~mm so that for the
thin regime, the condition $a \ll l_c$ is quite difficult to
experiment. The thin regime could be easier to attain if we could
increase $l_c$ and this may be achieved under microgravity conditions. Typical conditions of microgravity experiments lead to a value of $10^{-2}~g$ ($1~g=9.81$~m/s$^2$) in parabolic flys \cite{caravelle} ($l_c\simeq 17$~mm) and $10^{-6}~g$ in orbital stations \cite{orbital} ($l_c\simeq 1.7$~m). In order to make an experiment, one can choose a value of $g$ that enables to have a value of $l_c$ large enough but not
too much (the distance between the peaks is also proportional to $l_c$). Brand and Pettit made an experiment with a 8 mm deep layer on a parabolic fly \cite {pettit} (As the capillary length of their ferrofluid is $l_c=15.5$~mm in the low gravity phase of the parabolic fly, they were not actually in thin nor thick regime).  

\subsubsection{Thin-inertial film}

If the Reynolds number $Re\gg1$ i.e. $\frac{|s|a^2}{f}\gg1$, we obtain
the following equation\footnote{see \cite{note1} for the electric analogy}:

\begin{equation}
\label{eq}
s^2=a(-k^4+2\Phi k^3 \left(\frac{1+\mu_o/\mu}{2}\right)-k^2)
\end{equation}
Let us consider the function $c(\mu)=\frac{1+\mu_0/\mu}{2}$, where $c(\mu)\leq 1$.
At the onset of the instability, $\Phi =\frac{2}{1+\mu_o/\mu}$ ($\geq
1$) and $k_{crit}=1$. When $\Phi\geq\frac{2}{1+\mu_o/\mu}$, the wavenumber of maximum growth rate is given by:
$$
k_m=\frac{1}{4}(3\Phi c+\sqrt{9\Phi^2 c^2 -8})
$$
The asymptotic critical magnetic induction in the thin case is
different from the one of the thick case and always larger. In order
to quantify this difference, we have to solve implicit equations in
section (\ref{implicit}).\\
We denote $\varepsilon$ as $\varepsilon=\Phi c-1$. Near the
onset of instability, the growth rate can be developed, at lowest
order in $\varepsilon$ and $\delta k$, and leads to: 
$$
s^{2}(\Phi c=1+\varepsilon,k=k_{crit}+\delta k)=s^{2}(\varepsilon,\delta k)=a(2\varepsilon-\delta k^2)
$$ 
We obtain then that $s^2$  has the order of magnitude $a\varepsilon$.

The validity conditions are:
$$
\left\{
\begin{array}{lrc}
a\ll1& \mbox{(thin film)}\\
f^2/a^5\ll\varepsilon & \mbox{(inertial film)}
\end{array}
\right.
$$

\subsubsection{Thin-viscous film}
\label{mu}

When the Reynolds number $Re\ll1$ i.e. $\frac{|s|a^2}{f}\ll1$, the
equation (\ref{disprel}) becomes \cite{bac88}:
\begin{equation}
s=\frac{a^3}{3f}(-k^4+2\Phi k^3 (\frac{1+\mu_o/\mu}{2})-k^2)
\end{equation}
At the onset of the instability, $\Phi =\frac{2}{1+\mu_o/\mu}$ and $k_{crit}=1$.
When $\Phi\geq\frac{2}{1+\mu_o/\mu}$, the wavenumber of maximum growth rate is given by:
$$
k_m=\frac{1}{4}(3\Phi c+\sqrt{9\Phi^2 c^2-8})
$$
Near the onset, we develop the growth rate in $\varepsilon$ and
$\delta k$:
$$
s(\Phi c=1+\varepsilon,k=k_{crit}+\delta k)=(a^3/3f)(2\varepsilon-\delta k^2)
$$
or also $s$ of order of magnitude $(a^3/f)\varepsilon$.\\
The validity conditions are:
$$\left\{
\begin{array}{lrc}
a\ll1& \mbox{(thin film)}\\
f^2/a^5\gg\varepsilon & \mbox{(viscous film)}
\end{array}
\right.$$
The dispersion relations of the different cases are presented on
Figure \ref{dispa}.

\subsubsection{Near the onset of the instability}
In the thin case, we can notice that:\\
$\bullet$ At the onset of the instability, the thin film of
ferrofluid has a viscous behaviour.\\
$\bullet$ The cross-over to the inertial regime is reached when $\varepsilon= f^2/a^5$ (Figure \ref{dthin}).\\
In the case of a thin layer, $\frac{1}{a^5}$ tends to $0$: it results
in the fact that a thin film has a viscous behaviour, at the
onset and far above. With the ferrofluid EMG 507 ($f\simeq 0.0068$),
we reach the inertial regime when $H_0\simeq
H_{crit}^{thin}(1+\frac{\varepsilon}{2})=H_{crit}^{thin}(1+f^2/(2a^5))$
i.e. $H_0\simeq 3~H_{crit}^{thin}$ ($a=0.1$).

\section{Implicit equations for the asymptotic critical fields}
\label{implicit}

As the magnetic permeability depends on the magnetic field, each asymptotic value of the critical magnetic induction (thick or thin) satisfies an implicit equation.\\
$\bullet$ In the thick case, the onset is given by equations
(\ref{critepais}): it is then necessary to know
$\mu_r(H=H_{crit}^{thick})$ to calculate the critical values. In
calculations of section (\ref{equation}), the function $\mu_r(H)$ is a
constant equal to $\mu_r(H_{crit}^{thick})$ (since our linear approach
is valid near the onset). This leads to the implicit equation:
$H_{crit}^{thick}=f(\mu_r(H_{crit}^{thick})) \sqrt{2/\mu_0}(\rho g
\gamma)^{1/4}$  where $f(\mu_r)=\frac{\sqrt{1+1/\mu_r}}{\mu_r-1}$. The
function  $f(\mu_r)$ i.e. $H_{crit}^{thick}$  varies rapidly as a function of $\mu_r$.\\ 
$\bullet$ In the thin case, the implicit equation can be written:
\begin{equation}
\label{critmince}
H_{crit}^{thin}=
\sqrt{\frac{2}{1+ \frac{1}{\mu_r (H_{crit}^{thin})} }} ~ H_{crit}^{thick}
\end{equation}
where $\mu_r$ is the relative permeability of the ferrofluid. $H_{crit}^{thick}$ is determined by the equation (\ref{critepais}) but note that in equation (\ref{critmince}), $H_{crit}^{thick}$ is a constant.\\
The question arises as to when we can reach the asymptotic value
$\sqrt{2}$ of the ratio $H_{crit}^{thin}/H_{crit}^{thick}$ in order to
put easily in evidence experimentally the difference between the
asymptotic values of the critical magnetic inductions.
The Langevin's classical theory has been adapted to yield the superparamagnetic magnetization relationship between the applied field $H_0$ and the resultant magnetization $M$ of the particle collection. For a colloidal ferrofluid composed of particles of one size, we have:
\begin{equation}
\label{sat}
M=M_{sat}\left(\coth(\alpha)-\frac{1}{\alpha}\right)\equiv L(\alpha)
\end{equation}
whith $\alpha=\frac{mH_0}{\it{k}T}$, where $m$ is the magnitude of the
magnetic moment of a particule, $L$ is the Langevin function,
$M_{sat}$ the saturation moment of the ferrofluid and $\it{k}$ the
Boltzmann constant. When the initial permeability is appreciable, it is no longer permissible to neglect the interaction between the magnetic moments of the particles (\cite{shliomis}, \cite{kaiser}, \cite{bean}) and equation (\ref{sat}) has to be modified. We shall consider equation (\ref{sat}) as a good first approximation (while $\chi_i$ is not much larger than unity).
This leads to the expression of the relative magnetic permeability:
\begin{equation}
\mu_r(H_0)=1+\frac{M_{sat}}{H_0}L\left(3 \frac{H_0}{M_{sat}} \chi_i\right)
\end{equation}
This function of $H$ decreases from $\mu_r(0)=\chi_i +1$ to $1$.
With the ferrofluid APG 512~A ($\chi_i=1.4$), solving the
implicit equations (\ref{critepais}) and (\ref{critmince}) results in
$H_{crit}^{thick}=65.2$~Gauss and $H_{crit}^{thin}=77.0$~Gauss. The
ratio $H_{crit}^{thin}/H_{crit}^{thick}=1.18$ and the difference
$H_{crit}^{thin}-H_{crit}^{thick}=11.8$ Gauss. Table \ref{mutable}
presents other ferrofluids results: We note that the ratio
$H_{crit}^{thin}/H_{crit}^{thick}$ increases and the difference
$H_{crit}^{thin}-H_{crit}^{thick}$ decreases as $\chi_i$
increases. It is better to use a ferrofluid with a low value
of $\chi_i$ to put experimentally in evidence the difference between
$H_{crit}^{thin}$ and $H_{crit}^{thick}$.

\section{Conclusion}

We get the dispersion relation of a layer of magnetic fluid of any
thickness and viscosity, under a uniform vertical magnetic field. The critical
magnetic field and the critical wavenumber of this instability are found to be
thickness-dependent. The dispersion equation, simplified into four
asymptotic regimes, enables to explicit the expression of the critical
magnetic induction of a thin film of ferrofluid. Near the onset of the instability, we show that the behaviour of the
ferrofluid may be viscous or completely inertial; the behaviour which is manifested depends on the
characteristics of the ferrofluid (contained in the parameter $f$
proportional to $\nu$). In order to put in evidence experimentally the fact that the critical magnetic induction depends on the thickness, it is better to use a ferrofluid with a low value of the initial susceptiblity $\chi_i$.\\

\section{Acknowledgement}
Philippe Claudin is gratefully acknowledged for fruitful discussions. Ferrofluidics Corporation, and in particular Karim
Belgnaoui, are acknowledged for help in collecting physical data (Table \ref{data}) of the ferrofluids they propose.

\newpage

\bibliographystyle{unsrt}
\bibliography{ref}

\begin{thebibliography}{10}
\bibitem{note1}
When $\mu\to\infty$ in equation (\ref{eq}), we expect to find the corresponding
  dispersion equation of electric case. This is not right: we have to consider
  the equation $s=a(-k^4+2\phi k^3 A_1 -k^2)$ where
  $A_1=\frac{1+\mu_0/\mu}{1+\frac{\mu_0/\mu}{1+F(ka)}}$. We check that if
  $\mu\to \infty$, $A_1\to 1$ and we obtain the electrical case \cite{neron}.
  This equation leads to (\ref{eq}) after the following simplifications:
  $ka\ll1$, $\frac{ka}{\mu_0/\mu}\ll1$, so that we can't consider
  $\mu\to\infty$ in equation (\ref{eq}).
\end{thebibliography}

\newpage
\listoffigures

\newpage
\begin{figure}
\centerline{
\epsfig{file=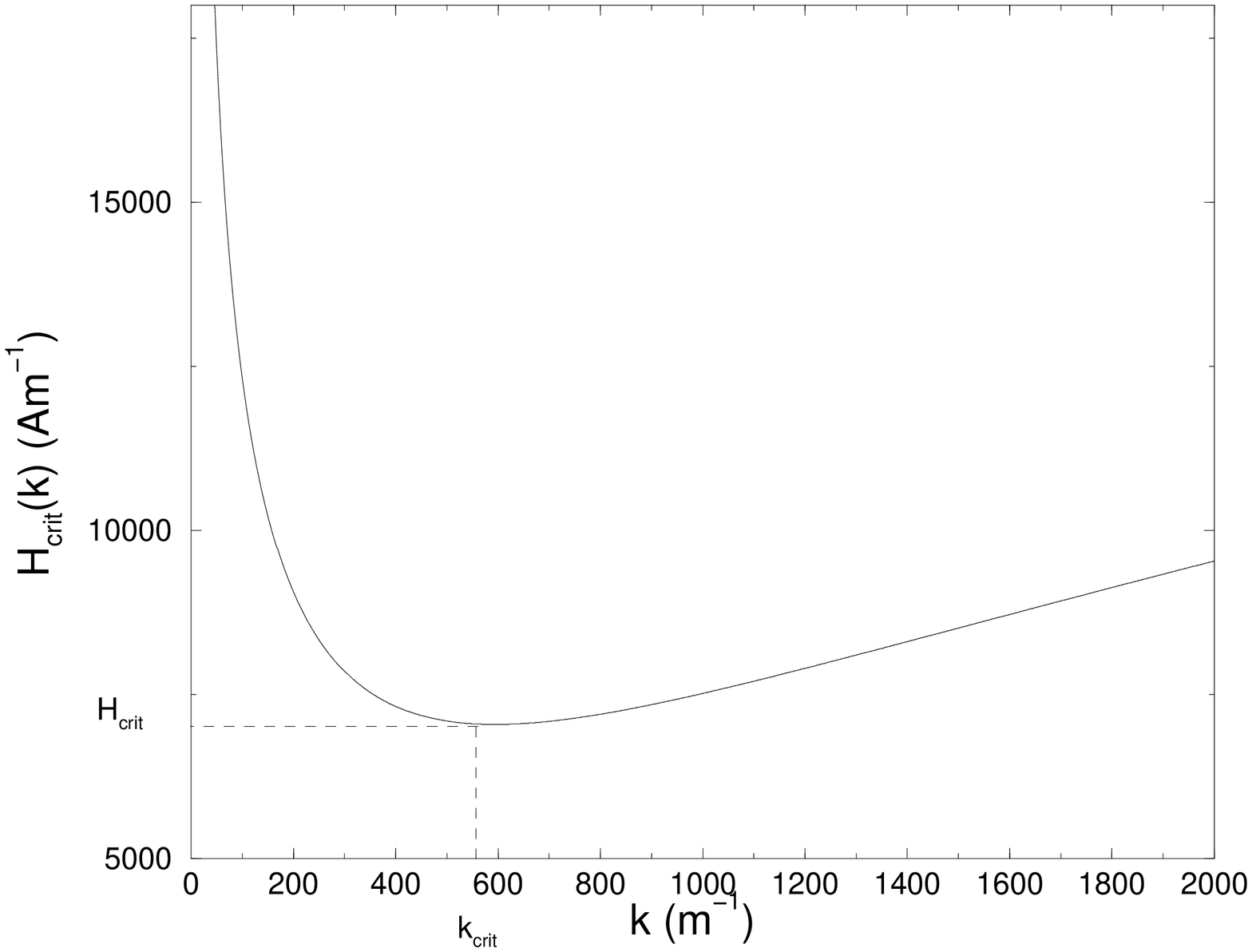,width=12cm,angle=-90}}
\caption{Curve of marginal stability; thick-inertial layer.}
\label{marginal}
\end{figure}

\begin{figure}
\centerline{
\epsfig{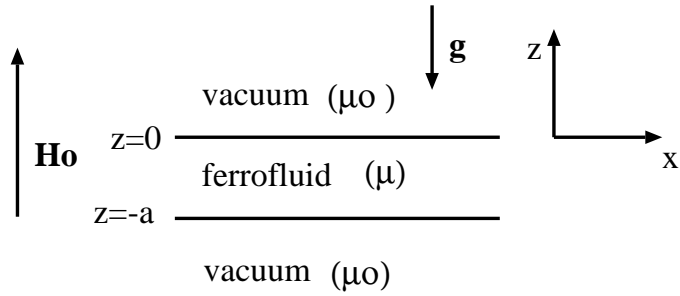}}
\caption{Ferrofluid layer in the vacuum and in a magnetic
  induction.}
\label{ferrovide}
\end{figure}

\newpage

\begin{figure}
\centerline{
\epsfig{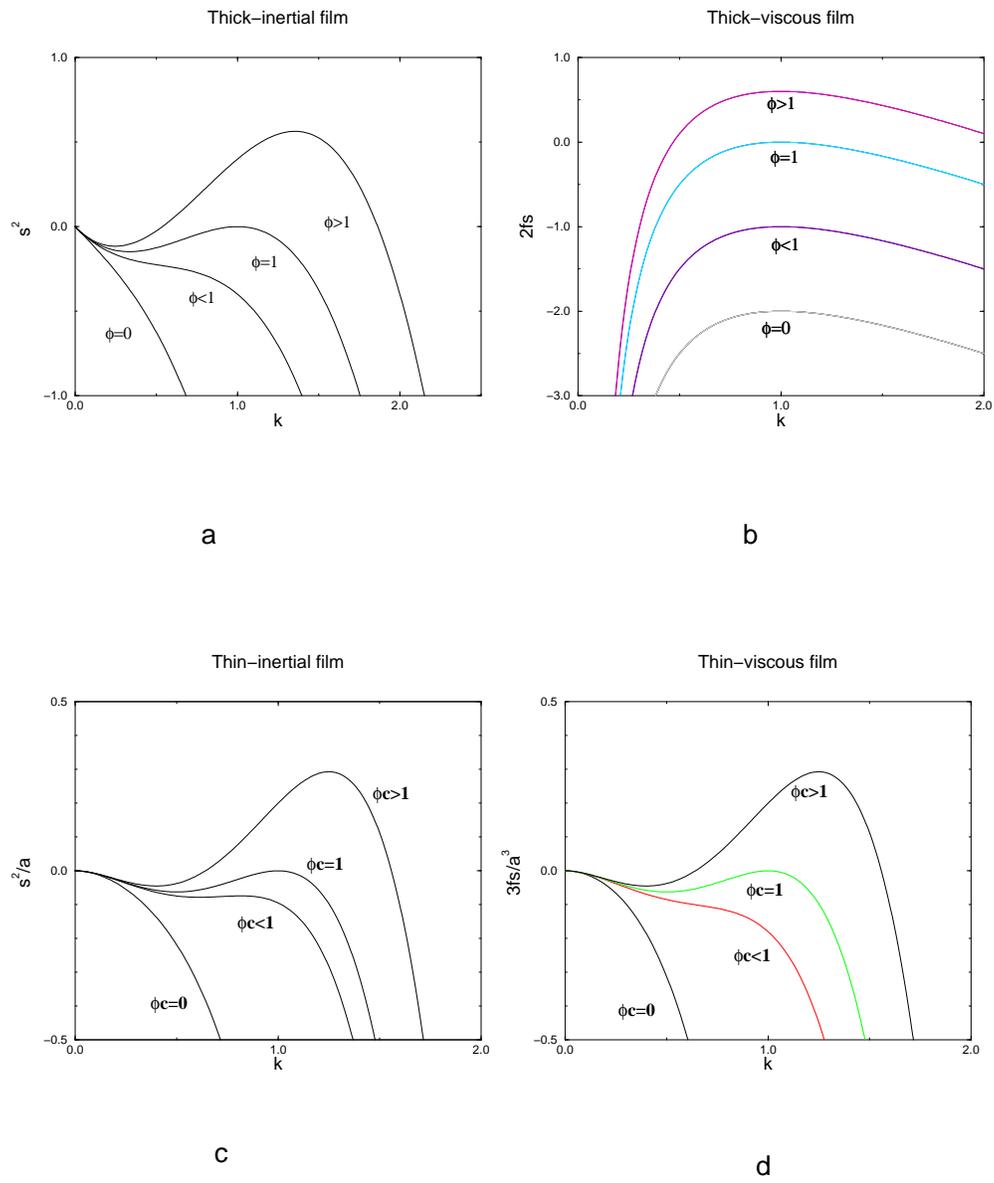}}
\caption{The dispersion relations:
${\bf a}$: $k_m$ depends on $\Phi$; $k_{crit}=1$;
$\Phi_{crit}=1$. ${\bf b}$: $k_m=k_{crit}=1$; $\Phi_{crit}=1$.
${\bf c}, {\bf d}$: $k_m$ depends on $\Phi$; $k_{crit}=1$;
$\Phi_{crit}=1/c=\frac{2}{1+\frac{\mu_0}{\mu}}$.}
\label{dispa}
\end{figure}

\begin{figure}
\centerline{
\epsfig{file=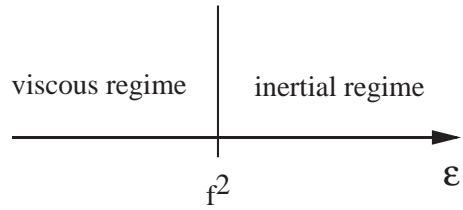,width=6cm}}
\caption{Viscous and inertial regimes, thick layer; $f=(l_v/l_c)^{3/2}$ and $\varepsilon=\Phi -1$.}
\label{dthick}
\end{figure}

\begin{figure}
\centerline{
\epsfig{file=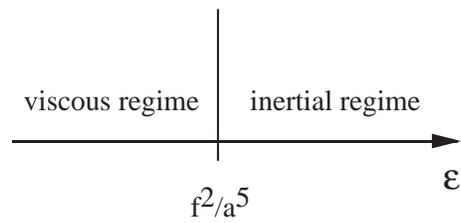,width=6cm}}
\caption{Viscous and inertial regimes, thin layer; $f=(l_v/l_c)^{3/2}$,
  $\varepsilon=\Phi c -1$ and $a$ is the layer thickness.}
\label{dthin}
\end{figure}

\clearpage
\listoftables

\newpage
\begin{table}
\begin{center}
\begin{tabular}{|c|c|c|}
\hline
       &  $\mbox{Re}\ll 1$  &  $\mbox{Re}\gg 1$  \\
\hline
$a/l_x\ll 1$ & thin-viscous regime: $\mbox{Re}=\frac{|s|a^2}{\nu}$ &
thin-inertial regime: $\mbox{Re}=\frac{|s|a^2}{\nu}$\\
\hline
$a/l_x\gg 1$ &  thick-viscous regime: $\mbox{Re}=\frac{|s|}{2\nu k^2}$ & thick-inertial regime: $\mbox{Re}=\frac{|s|}{2\nu k^2}$\\
\hline
\end{tabular}
\end{center}
\caption{The asymptotic regimes}
\label{tabregime}
\end{table}

\begin{table}
\begin{center}
\begin{tabular}{|c|c|}
\hline
capillary length            &  $l_c=(\gamma /\rho g)^{1/2}$\\
\hline
capillary time             &  $t_c=(l_c/g)^{1/2}$\\
\hline
capillary Laplace pressure  &  $p_c=\gamma/l_c$\\
\hline
\end{tabular}
\end{center}
\caption{Capillary quantities}
\label{tabcapillary}
\end{table}

\begin{table}
\begin{center}

\begin{tabular}{|c|c|c|}
\hline
ferrofluid & $f$ & $H_0/H_{crit}^{thick}$ when begins the inertial regime\\
\hline
EMG 507 & 0.0068 & 1.00002\\
\hline
EMG 901 & 0.039 & 1.0008\\
\hline
APG 512 A & 0.27 & 1.04\\ 
\hline
APG 314 & 0.71 & 1.25\\
\hline
APG 067 & 1.29 & 1.8\\
\hline 
\end{tabular}
\end{center}
\caption{Range of the viscous and inertial regimes in the thick case}
\label{thick-range}
\end{table}

\newpage

\begin{table}
\begin{center}
\begin{tabular}{|c|c|c|c|}
\hline
ferrofluid & EMG 308 & APG 512 A & EMG 900\\
\hline
$\chi_i$ & 0.3 & 1.4 & 4.2\\
\hline
$M_{sat}$ (Gauss)& 600 & 300 & 900\\
\hline
$H_{crit}^{thick}$ (Gauss) & 299.5 & 65.2 & 19.7\\
\hline 
$H_{crit}^{thin}$ (Gauss) & 318.2 & 77.0 & 25.5 \\
\hline 
$H_{crit}^{thin}/H_{crit}^{thick}$ & 1.06 & 1.18 & 1.29 \\
\hline
$H_{crit}^{thin}-H_{crit}^{thick}$ (Gauss) & 18.7 & 11.8 & 5.8\\ 
\hline
\end{tabular}
\end{center}
\caption{Critical characteristics of various ferrofluids}
\label{mutable}
\end{table}

\begin{table}
\begin{center}
\begin{tabular}{|c|c|c|c|c|}
\hline
ferrofluid & $\rho$ (g/cm$^3$) & $\gamma$ (N/m) & $\eta$ (mPa~s) & $\chi_i$\\
\hline
EMG 507 & 1.15 &  $\sim$ 0.033  & 2 & 0.4\\
\hline 
EMG 900 & 1.74 & 0.025  &  60 & 4.2\\
\hline
EMG 308 & 1.05 &  $\sim$ 0.04  & 5 & 0.3\\
\hline
EMG 901 & 1.53 & 0.0295 & 10 & 3\\
\hline
APG 512 A & 1.26 & 0.035 & 75 & 1.4\\ 
\hline
APG 314 & $\sim$ 1.2 & 0.025 & 150 & 1.2\\
\hline
APG 067 & 1.32 & 0.034 & 350 & 1.4\\
\hline
\end{tabular}
\end{center}
\caption{Physical data of ferrofluids (Ferrofluidics Corporation).}
\label{data}
\end{table}

\end{document}